\documentclass{acm_proc_article-sp}
\usepackage{graphicx}
\usepackage{epstopdf}
\usepackage{color}
\begin{document}

\title{GNU Radio Signal Processing Models for Dynamic Multi-User Burst Modems}

\numberofauthors{2} 

\author{
\alignauthor
Timothy J. O'Shea\\
       \affaddr{Virginia Tech Research Center}\\
       \affaddr{900 N Glebe Rd, Ste  200}\\
       \affaddr{Arlington, VA 22203}\\
       \email{oshea@vt.edu}
\alignauthor
Kiran Karra\\
       \affaddr{Virginia Tech Research Center}\\
       \affaddr{900 N Glebe Rd, Ste  200}\\
       \affaddr{Arlington, VA 22203}\\
       \email{kkarranc@vt.edu}}
\date{1 July 2015}
\maketitle

\begin{abstract}
This paper presents a modern method for implementing burst modems in GNU Radio.   Since burst modems are widely used for multi-user channel access and sharing in non-broadcast radio systems, this capability is critical to the development of numerous waveforms in GNU Radio.  We focus on making such systems easy to develop and adapt to wide classes of modems and computationally efficient at runtime.  We use the GNU Radio Event Stream scheduler to demonstrate concise implementations of burst PSK and FSK modems in GNU Radio and compare this with alternate approaches which have been attempted in GNU Radio.
\end{abstract}

\terms{GNU Radio, Software Radio, Architecture, SDR, Burst, Modem, Eventstream, Message Passing, Concurrency, Scheduler, PSK, FSK}


\section{Introduction}
GNU Radio \cite{gnuradio} has been through a long evolution of signal processing 
models and feature growth over its lifetime.  

GNU Radio's roots are in stream processing and as a consequence, most users of GNU Radio have remained application focused, making do with the systems in place to implement their wireless research applications of interest.   

For continuously modulated broadcast applications such as FM audio or television/MPEG stream broadcast, this stream processing architecture was exceedingly well suited.  As GNU Radio grew into the swiss army knife of software radio applications, people became increasingly interested in other applications including packet based and multi-user radio standards in which channel access needed to be carefully controlled and timed by each node.

Initially, building applications for such protocols often meant implementing monolithic and specialized signal processing blocks that combined many operations and external data structures into a single stream actor with sizable state machines and internal logic.

Today, GNU Radio has a number of architectural features which make implementing these complex bursty and stateful waveforms in clean modular ways significantly easier.  This push towards standardized subsystems for each signal processing model has increased block reuse opportunities, increased block generalization to many problems, decreased effort required to implement new blocks by leveraging these models, and generally empowered GNU Radio users and applications to tackle bigger problems than were possible before for the same sized research effort.  

This paper begins by reviewing the models available in GNU Radio to design reusable modems.  Next, we dive deeper into the burst model and provide reference designs for reusable burst modems for multi-user PSK and FSK communications systems.


\section{Signal Processing Models}

GNU Radio's key signal processing models are the stream processing and message processing models.  There are also several methods for representing non-contiguous streams of items which we will review before detailing how they may be used in conjunction.  As will become clear, the choice of which signal processing model one should use to build their modem depends on the application that one is designing their system to conform to.

\subsection{Stream Processing}
GNU Radio's origins lie in the stream processing model.  As stated above, the stream processing model works well for continuously modulated streams of data, such as broadcast FM.  In the stream processing model, every block in the waveform is "always on" and processing data.  

The key component in GNU Radio which enables stream processing is GNU Radio's stream scheduler.  GNU Radio's efficient stream scheduler provides a way to build a set of stream blocks or actors which consume and produce "items" between input and output circular buffers.  Items are generally intended to be small scalars such as u\_int8, float32, or complex64 data types, but have in some cases been used to hold larger "items' such as an entire C struct.  The scheduler in general decides how many items should be produced.  Thus, the stream scheduler enables stream blocks or actors to be chained together to create a waveform without having to worry about the details of how data flows between the blocks.

Although the stream processing model works extremely well for older continuously modulated streams of data, it's efficiency is questionable when trying to adapt it to waveforms that might require sideband signaling information along with the data stream.  In response, the stream processing model was evolved to include \textbf{Stream Tags}.

\subsection{Stream Processing Evolution}

Stream tags are a way to introduce out-of-band annotate arbitrary information onto the item stream.  
This was rapidly adopted for instance by the Ettus Research USRP's UHD driver to annotate "rx\_time", "rx\_rate",
and "rx\_freq" tags periodically onto a received sample stream to provide information about what the
sample stream represented to downstream blocks.   In general this mechanism replaced the need for blocks
to emit parallel signaling streams of zeros and ones to annotate timed information by providing a much more efficient means.

Once the idea of using stream tags for signaling information was established, the notion of \textbf{Tagged Stream Blocks}
emerged.  The idea was that by annotating chunks of stream items into packets or frames, the existing stream scheduler and blocks could be used to handle larger blocks of information atomically in a work function.
This is typically is done by annotating a "packet\_len" tag with an integer number of items 
on the first item on the stream, and on any item immediately following a previous "packet", but has also taken the form (such as in UHD) of "start-of-burst" and "end-of-burst" stream tags, which allows for longer chunks.
This provided a nice interoperable mechanism to pass chunks of items around in the existing scheduler, but runs into issues fitting large and variable chunk sizes in pre-allocated circular buffers.  As an alternative to using circular buffers for all communications, the message based processing model was introduced into GNU Radio. 

\subsection{Message Passing}

GNU Radio \textbf{Message Ports} were formally introduced in version 3.7, providing the ability for blocks to pass discrete messages around in addition to or instead of using the
traditional stream ports.   Message ports are named and follow a publisher/subscriber model where receive queues exist at message input ports.
A number of common message tasks such as Socket I/O, Tunnel/Tap I/O and conversion to and from Tagged Stream Block form are provided in tree.
Messages are defined as any GNU Radio \textbf{PMT (polymorphic type item)}, but some more well defined types exist.   The \textbf{PDU} or \textbf{protocol data unit} structures are the most notable of these which consisted of a Tuple containing a metadata dictionary and a uniform vector of samples (complex64, u\_int8, or otherwise).
Initial message based blocks used these PDU types as a standard message format, and this
has now been widely adopted for interoperability.
 
\subsection{Bridging Streams and Message Passing}

The prior two models, stream and message passing both provide powerful basic building blocks for many signal processing systems.   
Most signal processing systems fit well into a stream half and a message passing half, partitioned somewhere in the middle.
The remaining hurdle lies when \textbf{translating between these two domains, which turns out to be a common stumbling block}.   
Extracting stream items or inserting stream items to/from message passed chunks are not complex operations, but can be a tedious, error prone, and duplicitous when dealing with scheduler complexities or when attempting to re-implement as part of a complex monolithic special purpose stream block's state machine.
To address this problem, the GNU Radio Eventstream scheduler \cite{eventstream} \cite{vteventmodel} introduces two principal blocks.
\begin{itemize}
  \item The \textbf{Eventstream Sink}, consumes a stream port and emits discrete events/messages downstream 
  \item The \textbf{Eventstream Source}, consumes discrete events/messages and places them into an outgoing stream precisely or opportunistically
\end{itemize}

Eventstream provides the notion of a \textbf{Trigger Block} which instructs an event to be either placed in or extracted from a stream at a specific time or sample index.
A common trigger block is simply a threshold detector, which queues a stream event upon a detection metric rising over a threshold.

\subsection{Signal Processing in GNU Radio}

The stream processing model, message processing model, and eventstream allow for rapid construction of burst modems and are able to implement many varieties of complex multi-user channel access and precisely controlled slot timing schemes as demanded in a relatively straightforward manner.

In the sections that follow, we discuss a reference architectures for building stateful burst modems which maintains block generality, leverages both GNU Radio's stream and message passing domains where appropriate, and uses the eventstream scheduler to provide rapid translation between these domains with minimal, if any, waveform specific code.


\section{Burst Modems in GNU Radio}
\subsection{High level outline of a burst modem}

From an implementation perspective, burst modems designed in GNU Radio can be split into two different domains, the message based domain and the stream based domain.  The message domain consists typically of components that perform packet based processing, such as block encoding, framing, or block randomization.  On the contrary, the stream domain is typically closer to the front end and does not need to account for packet boundaries when processing data.  Examples of components in the stream domain include filtering operations, gain control, continuous tracking loops, and other streaming algorithms.  These two sections of the waveform are combined in order to make a burst transmitter or receiver through a translation layer.  In our work, we use the eventstream library to perform this translation.  

A transmitter burst waveform begins in the message domain, which generates packets that are to be transmitted.  These packets are first processed according to the MAC/PHY protocols and then sent to the streaming domain where final processing on the symbols and continuous waveform may be done before transmission.  A burst receiver waveform typically begins with a stream of sample items from an ADC, performs signal conditioning, gain control, filtering, and burst detection.  Following these stages the message domain processes each burst independently, performing block synchronization, demapping, decoding, descrambling, deframing, and other packet operations.

\begin{figure}[htp]
\centering
\includegraphics[scale=.3]{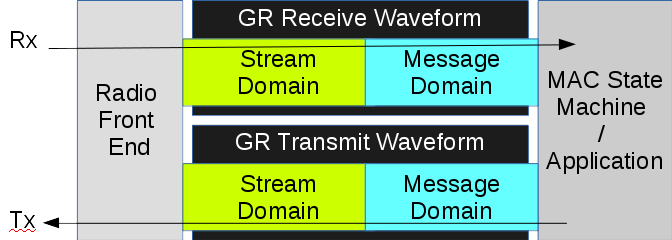}
\caption{General reference architecture}
\label{ref_arch}
\end{figure}

\subsubsection{Transmit Waveform Overview}

A typical transmit burst waveform will begin by first sourcing some PDU packet data possibly fragmenting it for the physical layer maximum transmission unit (MTU) if required.  This may involve fitting chunks of the network layer PDU into link layer PDUs, adding error checking, or additional signaling overhead.    Depending on the protocol, a block based FEC mechanism such as block turbo codes (BTC) or LDPC codes may be inserted to provide error correcting capabilities.  A block randomizer may be applied to whiten the resulting symbol stream.  Finally a preamble or reference signal is typically inserted to aid in receiver detection and synchronization; this will be discussed in further detail in the receiver section.  These are all typically message passing functions in the transmit waveform.  Next, the bits are typically mapped to symbols based on the modulation scheme, which may be the final message passing function.  At this stage, the eventstream source block may be used to schedule these symbols into an outgoing symbol rate sample stream.  After conversion to streaming, continuous functions such as pulse shaping, filtering, and transmission to a streaming software radio front-end device and DAC is typically conducted.

\subsubsection{Receive Waveform Overview}
The burst receive waveform is somewhat of an inverse of operations in the transmitter in reverse order.  Because the receiver must estimate timing, synchronization information and best mitigate channel noise and impairments, it typically includes additional algorithms to synchronize and perform optimal symbol and FEC codeword estimation.  The first step in the receiver is to digitize the data and reverse any pulse shaping that was applied in the transmitter, typically as stream operations.  After receive pulse shaping, an energy detector or waveform specific detection metric may then detect burst arrival.  This block serves as an eventstream "trigger" or a stream input to a standard threshold trigger, for the bridge between the streaming and message passing domains.   Based on trigger timing information an eventstream sink block may consume a stream and produce correctly timed message events to push into the message passing half of the waveform.

Once bursts of samples are detected and extracted, a block synchronizer may be employed to resolve time and frequency ambiguities.  In waveforms with continuous signaling, tracking loops may be used to perform this function, however, for burst architectures, we elect to employ block based synchronizers to avoid pull in time in time and and resulting corruption of symbols at the beginning of a burst.  Any residual equalization, rotation or symbol time offset errors may then typically be resolved by comparison with a preamble or reference signal.  Synchronized symbols may then demapped to bits either hard bits or soft log likelihood values. 

Packets of bits may then be passed through forward error correction decoding, de-randomization, de-framing and any other necessary packet processing operations to invert transmitter encoding before passing message and events to a high layer end application or MAC state machine.

\subsection{Reference PSK Burst Modem}

This section details the architecture of a QPSK burst modem designed and built using the message passing framework in GNURadio.  At a high level, it consists of transmitter and receiver waveforms which follows the concepts laid out above, from an architecture viewpoint.  The sections below detail the implementation of these waveforms.  

We diverge somewhat from the more conventional GNU Radio PSK modem in that, we use burst based estimates for synchronization estimation rather than tracking loops which take time to converge.   The correlate and sync block recently has attempted to address this for stream graphs, but we show purely message based synchronization approach here.
 
\subsubsection{Transmit Waveform Overview}
 
As noted above, the burst transmitter waveform begins in the message domain.  The first block in the message domain is the "Message Strobe" block, which generates a protocol data unit (PDU) periodically every 1000ms; this triggers the creation of a random data PDU corresponding to each of these.   These random data PDUs are simply a vector of bits with random length and an empty dictionary which could hold information about the burst.  In reality an application might implement a higher layer MAC here, or simply use a TUNTAP or SOCKET block to allow real data PDUs to flow into the graph, but this is a convenient model for testing.  These random bits are then passed into the framer block which adds a length field, a header checksum, and a payload checksum - which will allow us to verify correct reception and determine packet length upon receipt of the burst at the downstream receiver.

After the framer, the PDU is passed through a "burst padder" block which simply adds zero bits to the end of the frame until the length reaches an integer multiple of the uncoded forward error correction (FEC) block size.   Because FEC blocks must be encoded in single multiples of the block size, this is a required step.  When operating without FEC this could be easily removed.   The PDU's are then sent through a randomizer which XORs a random sequence onto the data to whiten the payload bits and pass it through the FEC encoder.

Finally, a known binary preamble sequence is added onto the front of each burst, and pass the burst bits through a QPSK bit to symbol mapper which converts the uint8\_t bit vector into a complex float32 sample vector ready to be inserted into our transmission stream.

The last block in the message domain for the transmit waveform is the burst scheduler, which decides when in time to schedule the burst.   It is simply a slotted aloha type of scheduler (although any kind of scheduling can be used) which drops the burst into the stream as soon as it can along some fixed slot boundary.   It sends bursts annotated with a sample-time to schedule them into the stream on to the eventstream source block, and receives asynchronous feedback from the same eventstream source block letting it know where in the stream it is "now".   Out of this eventstream source block, we get a sample of complex zero samples with events copped into the correct offsets specified by their event time.

The stream domain for this transmitter consists of three blocks, a throttle block to limit the speed of the outgoing sample stream, an interpolating RRC filter to convert from 1 sample per symbol up to 2 samples per symbol for transmission, and a standard channel model block are used to simulate transmission.  The output of the channel model block is then connected to the a file sink and some QtGui plotters.

The image below shows the total transmit flow-graph as described above.

\begin{figure}[htp]
\centering
\includegraphics[scale=.45]{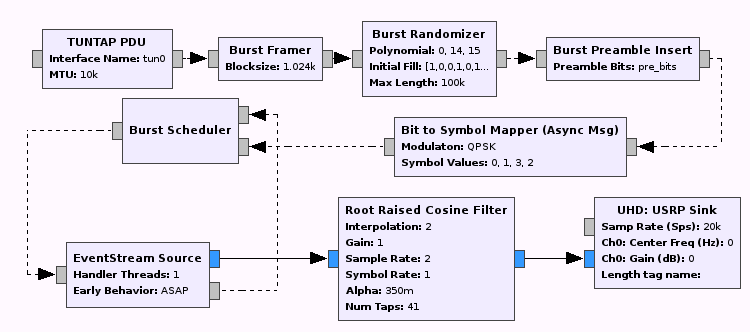}
\caption{PSK Burst Transmitter}
\label{burst_tx}
\end{figure}

\subsubsection{Receive Waveform Overview}

As noted in the general section above, the receive waveform begins in the streaming domain with a digitizer and receive filtering.  A sample stream, either from a radio or out of a stored sample file begins the processing.   To simulate real time operation, we throttle the data stream at the intended sample rate and then run through a matched filter for the presence of a preamble.   The output of this filter is then further filtered through a local comparison to a moving average, in the correlator filter block, and then the resulting detection metric is run into the eventstream trigger rising edge block.  This block detects whenever the detection metric rises over a certain threshold at the beginning of a burst and sends an asynchronous message to the eventstream sink block to extract a burst event beginning at that time in the sample stream.  As stated above, this eventstream block is the bridge between the streaming and message domains in the receiver waveform.

The message based section of the received waveform begins with the chunk of samples extracted from the continuous stream containing the burst somewhere in it.   Fine synchronization has not yet been completed, and the length of the burst is not yet known exactly, so an upper bound on the length of all bursts worth of samples is extracted.  This is immediately plotted as power over time, and then run through a length detector block which attempts to trim some of the noise off the end of the burst based on its power envelope fall off / trailing edge.

Having hopefully minimized the number of samples, the next step is to run through the synchronization algorithm, which is of non-trivial compute complexity.  The synchronizer produces a maximum likelihood estimate for the CFO over the length of the burst, computes optimal timing and equalizer taps over the length of the burst, and then applies them automatically to the entire burst worth of samples.    The lower left hand plot below shows the correlation peak obtained during timing synchronization within this block,  with a clearly observable peak at the preamble.

The output of the synchronizer block is burst symbols lined up in the correct rotation based on the preamble.  The next block, the soft demapper translates all complex float symbols in the burst into a series of floating point soft bits in a PDU.   The burst frame align block then strips the preamble bits off the front of the burst, and ensures that all the remaining bits are trimmed to a multiple of the coded FEC block size (in this case 271 bits).  These soft bit vectors are then passed through the FEC decoder block and a derandomizer block to perform LDPC decoding and output hard bits and remove the XORed whitening sequence.

The decoded, derandomized hard bits are then sent through a deframer which computes a CRC and removes any extraneous trailing zeros from the packet.  The resulting PDU is then sent to a "meta text output" GUI widget, which allows us to look at the PDU's dictionary values for each burst in a nice clean and easy way for visualization.

\begin{figure}[htp]
\centering
\includegraphics[scale=.35]{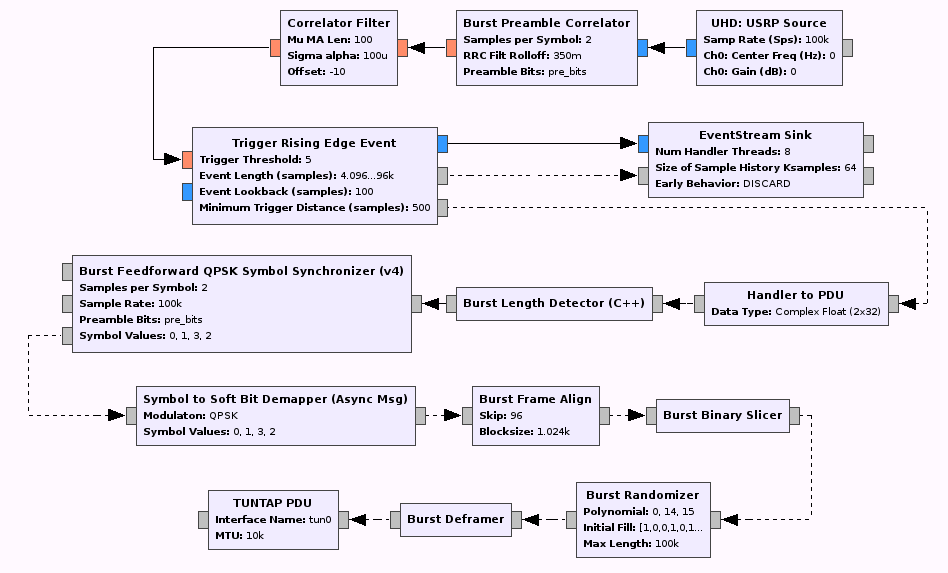}
\caption{Burst PSK Receiver}
\label{burst_rx}
\end{figure}

\clearpage
\subsection{Reference FSK Burst Modem}

Burst Frequency Shift Keying, FSK, modems are widely popular in low cost commercial and household electronics due to their simplicity and cost to implement.  Many RFID systems, garage door openers, automotive remote lock systems, and low cost sensor systems use burst FSK modems to transmit information.
GNU Radio has long been able to decode FSK burst transmissions, but until recently this involved simply running continuous tracking loops for frequency tracking, and timing recovery and using a simple binary matched filter known as a "correlate access code" block to detect starts of bursts.   This approach worked, but was both computationally wasteful and required a fairly specialized correlate access code block which understood symbol mapping as well as stateful protocol information.
We propose an alternative in which we still continuously run the frequency tracking loop (quadrature demodulator) in the stream domain, but move timing recovery, equalization, start of burst alignment, and all other operations into the message passing domain. 

\subsubsection{Transmit Waveform Overview}

FSK Transmission is relatively straightforward, the message passing portion of the waveform often appears similar to that of the previously described PSK transmitter, except many low cost systems use little or no forward error correction, scrambling, or complex framing schemes.  This means transmit waveforms are often relatively simple and do not include complex packet operations up front.  Mapping to symbols is then the primary differentiator where instead of mapping to a complex amplitude and phase value at baseband, an integer deviation value, typically +1,-1 mapping for two-level FSK, is mapped and then mixed with a frequency modulator or voltage controlled oscillator over the burst time window.
Scheduling the burst into the stream and performing any additional stream conditioning filters or resampling processes is then conducted in the same way as the PSK modem resulting in a continuous stream of samples to be transmitted to the DAC for upconversion to pass band and transmission through your favorite radio front end.

\begin{figure}[htp]
\centering
\includegraphics[scale=.4]{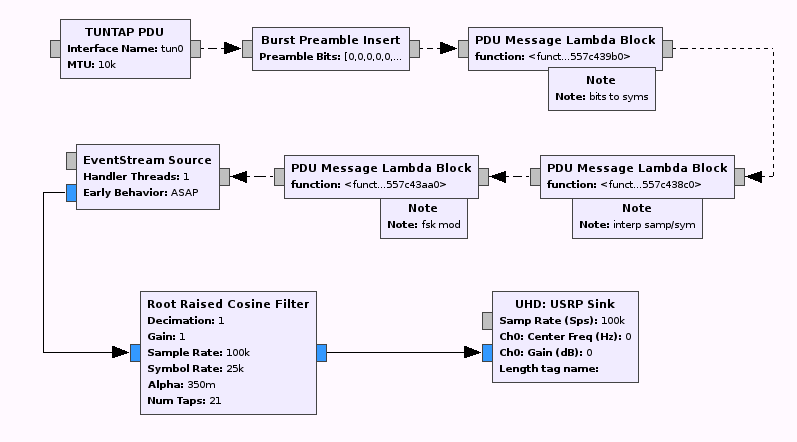}
\caption{FSK Burst Transmitter}
\label{burst_fsk_tx}
\end{figure}

As a side note, several of the message blocks in the transmit modem pictured are message lambda blocks.   These message lambda blocks allow for the rapid creation of new blocks by defining a single python function which maps input to output messages.  This is a useful prototyping tool which allows for rapid creation of new blocks from within GRC while developing prototype waveforms.

\subsubsection{Receive Waveform Overview}

The receive waveform for a burst FSK receiver typically begins with a sample stream from an ADC and SDR front-end device, followed by some amount of streaming signal conditioning, filtering, gain control, and/or sample rate conversion.   Energy detection could be used as a metric before any form of synchronization, but we have opted instead to at least perform the frequency tracking loop continuously in the frequency domain, because the variance of the continuous FM demod signal actually doubles as an effective burst detection metric, decreasing greatly in variance when locked to an FSK carrier.
We use a standard threshold trigger block to queue event extraction when FM demod variance falls below a threshold value, then pulling a burst worth of FM demodulated samples into the message domain for the remainder of the modem.
Timing recovery and preamble synchronization are performed only when messages arrive after detection on a window of burst samples, leading to gains in processing efficiency for low duty cycle systems.   Symbols may then be equalized and sliced according to the FSK modulation order, and then handled as packets of bits through decoding, descrambling, deframing, etc in a method identical to that of the PSK burst receiver.
Finally the decoded output PDUs may be passed to a higher level MAC state machine or external application layer through a socket or a tunnel/tap interface with the operating system.

\begin{figure}[htp]
\centering
\includegraphics[scale=.37]{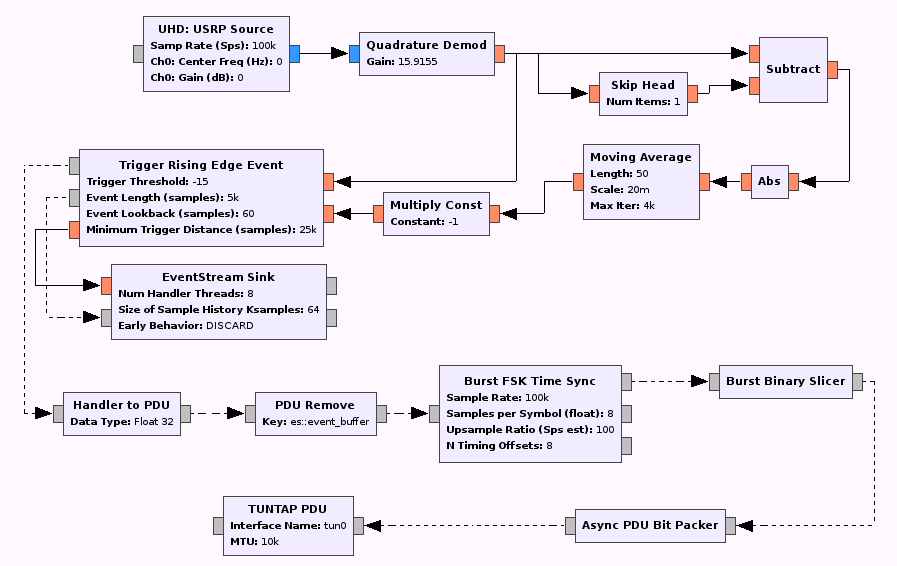}
\caption{FSK Burst Receiver}
\label{burst_fsk_rx}
\end{figure}

\subsection{ACM Feedback Problem}

Numerous modern wireless protocols including WiFi have adopted \textbf{Adaptive Coding and Modulation} schemes.  This means that the modem's error correction code rate and modulation order are matched to the track the information capacity of the wireless channel as it varies to maximize usable spectral efficiency while accessing the channel.
Using a pure stream proccessing model, GNU Radio was constrained to forward-only flowgraphs or "directed acyclic graphs", DAGs, which disallowed the communication of adaptation information back upstream in a flow graph, for instance from a header decoder to control a demodulator or decoder block.
By using message passing to conduct upstream control signalling or pure message domain signalling and extraction of ACM payload information, this is now a solved problem.

The GNU Radio OFDM modem \cite{ofdmmodem} \cite{grofdmfosdem} for instance uses a tagged stream demux block to receive decoded control information from downstream while then extracting and passing data and control to a downstream payload region demodulator.  This one great solution for how to do reconfiguration, but it suffers from a lack of concurrency and from a long dependency loop which must wait for header demodulation to extract a payload before extracting the next header.

An alternative approach using the event-stream scheduler to decode a header and then queue a secondary payload event in comparison, need not block tightly on the consumption of the payload samples.   Allowing more work to be done concurrently to maximize protocol throughput.

In either architecture, GNU Radio is now equipped to deal with complex adaptive coding and modulation schemes used in modern wireless communications signals while maintaining algorithm generality, module re-use and processing efficiency.

\subsection{Burst Plotting Tools}

One issue which quickly comes to light when developing message and event based modems is the need to plot discrete events and messages for diagnostic purposes.  Unfortunately GNU Radio's current graphical plotting widgets, gr-qtgui and gr-wxgui, were designed for stream processing only.  
As message based modems and applications become more prevelent, plotting and diagnosis tools intended for use with complex and real valued message PDUs are needed to fill this gap.
Ultimately, with the right message port interface a unified set of in-tree supported unified plotting tools is likely the right solution for GNU RAdio, but as a short term interim solution we provide the gr-pyqt module, a set of message based plotters for GNU Radio PDU formatted message types.   An example of a diagnostic GUI for a QPSK modem using these gr-pyqt based message plotters is shown in the image below, and a file plotting application driven entirely by message passing is included in the module called "Such Samples."

\begin{figure}[htp]
\centering
\includegraphics[scale=.24]{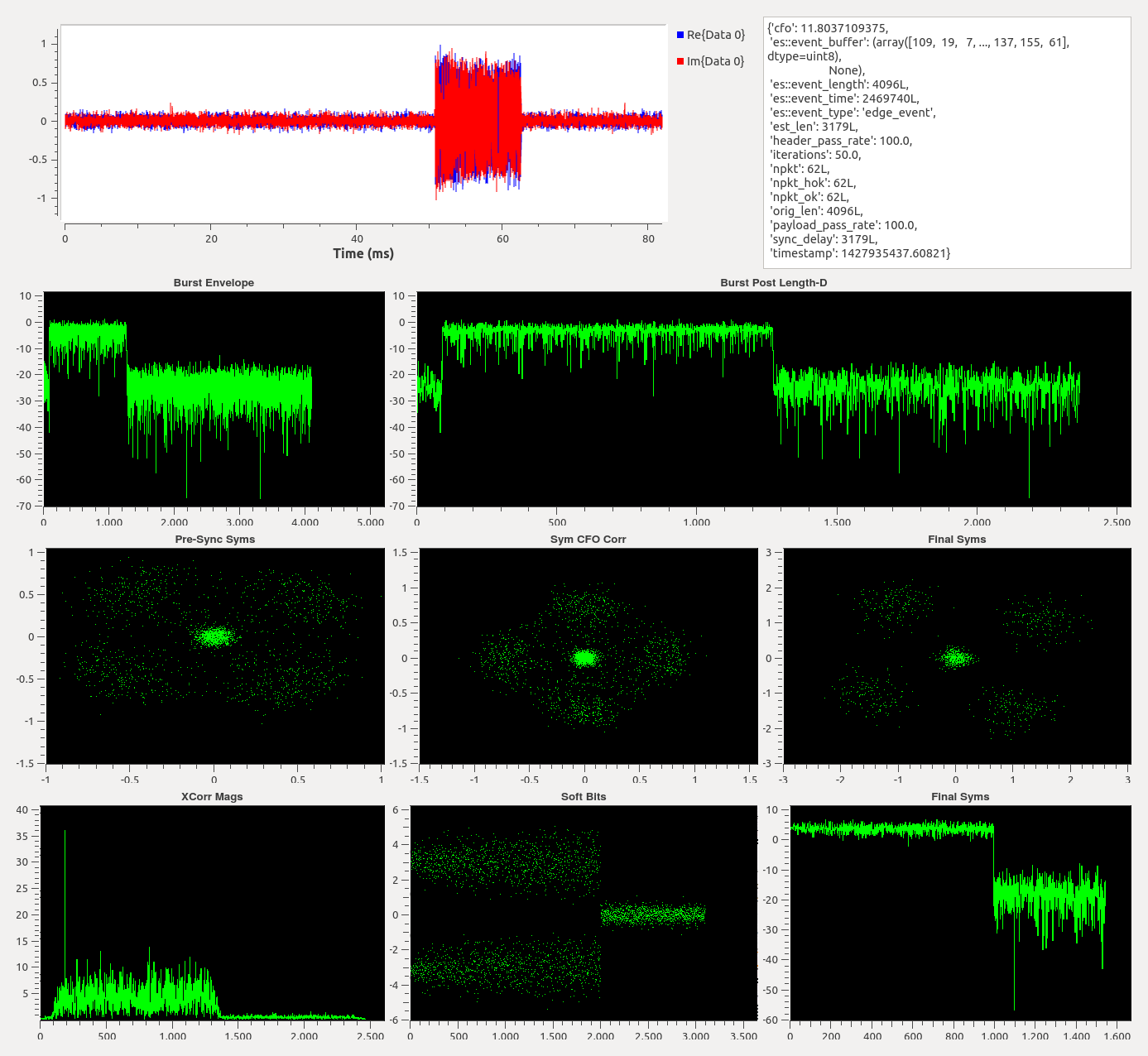}
\caption{Message Based Plotting with GR-PyQT}
\label{rx_waveform}
\end{figure}


\section{Conclusions}

In this paper, we have presented reference architectures for bursty multi-user signal processing in GNU Radio.  GNU Radio signal processing models including stream and message passing, along with conversion between them were explained.  While pure stream based applications were sufficient for some legacy broadcast applications, the motivation for tagged stream and message passing models is the need for shared and well controlled wireless channel access and waveform adaptation.

By presenting a general reference architecture along with examples of both PSK and FSK burst modems, it is our hope that we can ease and standardize the development of more mature, modern, capable, and wireless standards interoperable modems by future GNU Radio users.
Through proper use of each of these signal processing models, high levels of code re-use, portability and interoperability can be achieved allowing for easier, faster and more intuitive modem design and testing.  
\\
\section{Future Work}

While GNU Radio has come a long way in its support for domain and application appropriate architecture and tools, there are still a number of areas for improvement and future work.
As identified earlier, better unified graphical tools for plotting and visual diagnostics of hybrid stream/message systems would be a great help.
Further benchmarking and performance comparison of both stream and message passing systems, schedulers and data structures could provide increased waveform throughput in numerous applications.   
Message passing models could in the future fairly easily relax in-order assumptions for state free message passing blocks, allowing features such as concurrent dispatch of multiple messages in a single message queue through the message passing scheduler.   This could provide significant throughput gains on many-core processing architectures without increased application complexity.
Portability of algorithms between block type, be they stream, message, tagged stream or otherwise is another major area for improvement.   As of the writing of this paper, blocks are typically written natively as one of these types, but in the future, an algorithm base class allowing for a single implementation of an algorithm followed by instantiation as one of the appropraite derived block processing model types could be created to further improve block library interoperability and code re-use.
Lastly, but critically, more mature example waveforms are needed, leveraging and making use of all of the modern architectural features of GNU Radio efficiently to serve as best practice exemplars and tutorials for new and advanced GNU Radio users to build upon and expand the ever growing toolset of the GNU Radio waveform ecosystem.

\nocite{gnuradiowiki}

\newpage
\bibliography{references}{}
\bibliographystyle{plain}

\end{document}